\documentclass[Journal,letterpaper]{ascelike-new}
\usepackage[utf8]{inputenc}
\usepackage[T1]{fontenc}
\usepackage{lmodern}
\usepackage{graphicx}
\usepackage[style=base,figurename=Fig.,labelfont=bf,labelsep=period]{caption}
\usepackage{subcaption}
\usepackage{amsmath}
\usepackage{algorithm,algorithmic}
\usepackage{newtxtext,newtxmath}
\usepackage[colorlinks=true,citecolor=red,linkcolor=black]{hyperref}
\NameTag{Sadasivan, \today}
\begin{document}
\nolinenumbers

\title{A Computational Approach for Multi-Body Potential-Flow Interaction Effects Using Matrix-Free FEM and Body-Conforming Grids}

\author[1]{Anil Lal Sadasivan}
\author[2]{Mannu Yadav}

\affil[1]{Associate Professor, Department of Mechanical Engineering, National Institute of Technology Sikkim,Corresponding author email: anillal65@gmail.com}
\affil[2]{Research Scholar,  Department of Mechanical Engineering, National Institute of Technology Sikkim, mannuyadavh@gmail.com}

\maketitle

\begin{abstract}
This paper presents a unified and computationally efficient framework for
predicting incompressible, irrotational (potential) flow around multiple
immersed bodies in two-dimensional domains, with particular emphasis on
quantifying irrotational interaction effects in multi-body configurations. The
methodology integrates three components: a fast body-conforming mesh-generation
strategy, a matrix-free finite-element solution of the Laplace equation, and a
systematic procedure for determining the stream-function values associated with
each immersed solid.

Body-fitted grids are generated by imposing boundary displacements on a
Cartesian background mesh followed by Laplacian smoothing, yielding simple,
robust, and accurate meshes for domains containing multiple immersed bodies.
The potential-flow field is obtained by solving the Laplace equation using
a matrix-free Conjugate Gradient method, wherein element-level operators are
evaluated without assembling global stiffness matrices. Immersed
bodies are treated as constant-\,$\psi$ streamlines, and their unknown
stream-function values are determined through multi-point constraints that
naturally capture inter-body flow connectivity.

To demonstrate the capabilities of the proposed framework, potential-flow
interference among five cylinders arranged in both streamwise and transverse
directions is examined in detail. Variations in the gap ratio~$\delta$ between
transverse cylinders produce changes in flow redistribution, stagnation
locations, and surface-velocity behaviour. The flow rate between the transverse
cylinders is shown to decrease significantly at small~$\delta$ because of
mutual irrotational interaction with the streamwise cylinders, and this effect
is quantified through the \textit{Coefficient of Potential Flow Interaction}
(CoPFI). The results highlight the ability of the proposed approach to resolve
subtle multi-body interaction phenomena with minimal case-setup effort and very
low memory requirements, while providing a quantitative measure of potential-flow
interference in complex immersed-body systems.
\end{abstract}

\section{Practical Applications}

The computational framework developed in this work is intended for rapid aerodynamic
evaluation in engineering settings where the interaction between multiple immersed
bodies plays an important role in system performance. Owing to the low computational
cost of the matrix-free finite-element formulation and the efficiency of the
body-conforming mesh generator, the methodology is well suited for applications that
require repeated evaluations of irrotational flow fields, parametric variations, or
design-space exploration.

In aerospace configurations, potential-flow interactions arise in several contexts.
The ability to generate body-fitted meshes rapidly and to
determine immersed-body stream-function values automatically makes the proposed
approach particularly attractive for conceptual and preliminary design stages, where
numerous geometric variants must be evaluated to understand mutual aerodynamic
influence.

Beyond aerospace, the framework is applicable to marine-hydrodynamic systems such as
closely spaced hulls, strut assemblies, hydrofoil clusters, and underwater vehicles 
operating in proximity. Industrial applications include flow analyses around tube
banks, heat-exchanger elements, and arrays of cylindrical or bluff bodies. In these
contexts, potential-flow models provide reliable first-order predictions of flow
deflection, local velocity amplification, and interference-induced loading trends.

The introduction of a Coefficient of Potential Flow Interaction (CoPFI) 
enables systematic quantification of interference effects in multi-body arrangements.
This metric may be incorporated into optimization procedures, surrogate models, or
reduced-order aerodynamic frameworks to guide decisions on spacing, orientation, and
overall layout of interacting components. As such, the proposed computational strategy
offers both methodological value and direct relevance to engineering design workflows
across the aerospace, marine, and industrial sectors.

\section{Introduction}

Low-fidelity fluid--mechanics models such as potA Computational Approach for Multi-Body Potential-Flow Interaction Effects Using Matrix-Free FEM and Body-Conforming Gridsential-flow solvers and simplified
body-conforming grid techniques offer substantial advantages for rapid aerodynamic
and hydrodynamic analysis, design exploration, and parametric studies.
Their computational efficiency enables fast evaluation of large design spaces and
supports sensitivity analysis and optimization procedures. Owing to the linear
elliptic nature of the governing equations, potential-flow formulations exhibit
smooth solution behaviour, numerical robustness, and ease of integration into
iterative design workflows . On the other hand, the solution of the Navier--Stokes
equations is computationally expensive because of the nonlinearity and the coupled
nature of the governing equations \cite{ferziger2002computational,fletcher1991computational,sherwin2005spectral}.

Potential-flow modelling using stream functions is widely employed for aerodynamic
analysis, hydrodynamics, and preliminary evaluation of flows around immersed bodies. The interaction between
multiple immersed bodies in fluids provides vital information for understanding and
designing systems that involve such configurations. When multiple bodies are present,
and when several design choices must be evaluated as part of an optimization
procedure, the numerical scheme to be adopted must enable
(i)~rapid development of body-fitted computational grids,
(ii)~robust solution of the Laplace equation, and
(iii)~accurate evaluation of streamlines, body stream functions, surface velocities,
and inter-body flow topology. Design optimization typically requires repeated
evaluations of the flow solution, which is extremely difficult to perform using
high-fidelity models such as the Navier--Stokes equations; therefore, low-fidelity
potential-flow models are attractive for optimization-focused studies.  The setup of a CFD
analysis framework is often cumbersome because of geometry modelling and mesh
generation \cite{gs2018inverse,abraham2021optimization,abraham2022}.

The present work focuses on developing three aspects that together form a unified
computational framework for potential-flow analysis around multiple immersed bodies:
(i)~a fast body-conforming mesh-generation procedure,
(ii)~a matrix-free finite-element solver for the Laplace equations governing the
stream function~$\psi$ and the velocity potential~$\phi$, and
(iii)~a systematic procedure for determining the stream-function values corresponding
to all immersed solid objects.

Classical mesh-generation strategies such as elliptic, parabolic, and hyperbolic
grid generation \cite{thompson1977grid,thompson1985survey,lal2001geometry},
algebraic smoothing \cite{eiseman1969smoothing}, Winslow smoothing
\cite{knupp1999winslow}, or unstructured triangulations
\cite{mavriplis1997unstructured} may become cumbersome or computationally intensive
for domains containing immersed objects. To address this, the present paper introduces
a simple and robust body-conforming grid-generation scheme. We show that the technique
is computationally inexpensive and effective for configurations involving multiple
immersed bodies. The current mesh produces mixed cells of triangular and quadrilateral nature and Finite Element Method (FEM) offers flexibility in dealing with these meshes and allows the easy implementation of boundary conditions \cite{zienkiewicz2005finite,donea2003finite,lal2009hybrid}.  

The second major contribution of this work is a matrix-free finite-element formulation
for solving the Laplace equations corresponding to the stream function~$\psi$ and
scalar potential~$\phi$ in incompressible, irrotational flows. The discretized Laplace operator results
in a symmetric linear system, which is solved using the Conjugate Gradient (CG) method
\cite{saad2003iterative}. A matrix-free implementation is employed: no
global stiffness matrix is assembled; instead, element-level contributions are
evaluated during each CG iteration. This significantly reduces memory
usage and enables the solution of large problems with many degrees of freedom
\cite{elman2005finite}. In the stream-function formulation, each immersed body
corresponds to a streamline and is therefore constrained to take a constant, but
\emph{a priori} unknown, value of~$\psi$
\cite{jiang1996evaluation}. Thus, as the
third component of our contribution, we incorporate multi-point constraints
individually on the surfaces of all immersed solids, enabling the evaluation of the
stream-function values.

The methodology is demonstrated on potential-flow configurations involving multiple
immersed cylinders. The flow configuration in the example problem include cylinders aligned along the $x$-axis as well as
additional cylinders positioned transversely. The interaction between neighbouring
bodies is analysed using streamlines and velocity fields. The results confirm that
the flow topology, stagnation-point locations, and rear stagnation behaviour emerge
naturally from the solution without explicitly enforcing the Kutta condition,
consistent with classical potential-flow theory
\cite{katz2001low,trefethen1980potential}. The resulting body surface stream function, flow nets and
tangential-velocity distributions demonstrate the accuracy and utility of both the
proposed grid-generation technique and the associated matrix-free potential-flow
solver. Along with describing different features of the flow, a Coefficient of
Potential Flow Interaction (CoPFI) has been reported as a function of the gap
ratio~$\delta$, which quantifies the extent of incompressible, irrotational interaction effects.

\section{Body-Conforming Grid Generation}

We consider a computational domain $\Omega$ covered by a uniform Cartesian 
background grid that encloses a set of immersed bodies. Each immersed body is 
represented by a closed boundary curve $\Gamma_i$, $i = 1,2,\dots,nib$, where 
$nib$ denotes the total number of immersed bodies. To enable fast detection of 
intersection points and to efficiently compute node splitting (defined below) and boundary 
displacements, we first identify the square sub-domains of the background grid 
that enclose each immersed boundary.

Next, the grid points lying immediately inside an immersed boundary are shifted 
onto the corresponding body surface. In certain cases, a single interior grid 
point may intersect the geometry such that it projects onto two distinct 
locations on the boundary; this situation is treated as a \emph{node-splitting 
event}. After projection, each point on the boundary is assigned its 
corresponding $x$- and $y$-components of the displacement.

For each such grid point $\boldsymbol{x}_{gk}$ located just inside the \(i^{\text{th}}\) body neighbouring its \(k^{\text{th}}\) point, we 
define the displacement vector  
\[
    \boldsymbol{d}_{ik} = \boldsymbol{x}_{ik} - \boldsymbol{x}_{gk},
\]
where $\boldsymbol{x}_{ik}$ is position vector of the associated point on the immersed boundary 
$\Gamma_i$. Here, $k$ indexes the individual points on $\Gamma_i$. To ensure 
smoothness in the vicinity of the boundary, these displacements are propagated 
into the surrounding domain by solving  
\[
\nabla^2 \boldsymbol{d} = 0, 
\qquad 
\boldsymbol{d}\big|_{\Gamma_i} = \boldsymbol{d}_i, 
\qquad 
\boldsymbol{d}\big|_{\Gamma_j} = 0,
\]
where the subscript $j$ refers to those boundaries whose nodes are aligned 
with the Cartesian background grid. The vertices of the final mesh are updated by 
\[
\boldsymbol{x}_{\text{new}} = 
\boldsymbol{x}_{\text{old}} + \boldsymbol{d}
\]

\begin{figure}[h]
\begin{center}
\includegraphics[width=0.80\textwidth]{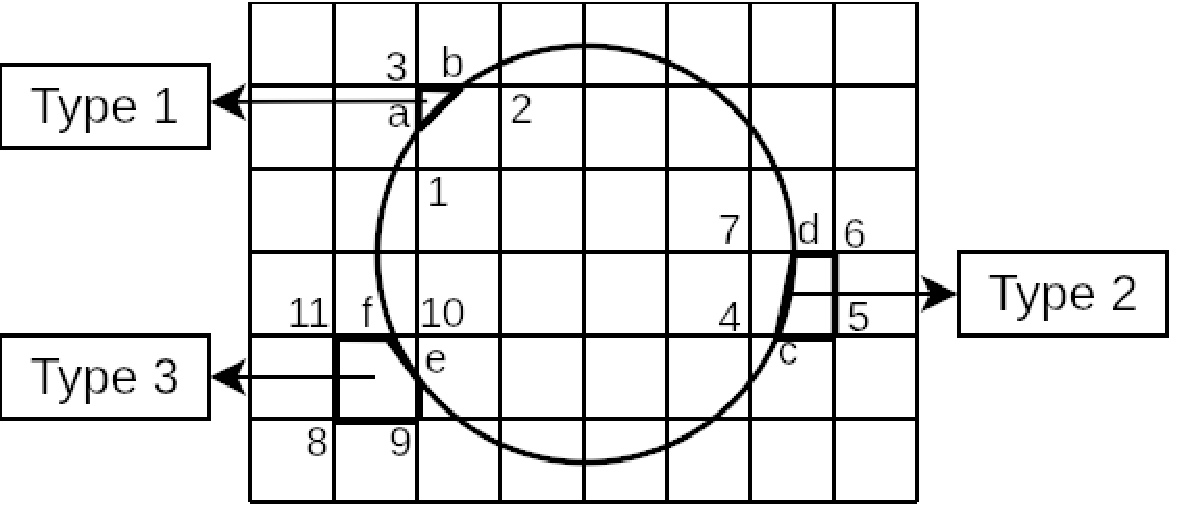}
\end{center}
\caption{Schematic diagram  of the mixed-cells mesh formation over an immersed body.}
\label{fig:1}
\end{figure}

Figure~\ref{fig:1} shows the resulting grid around a circular geometry contained 
within a background grid sub-domain. Node-splitting leads to modifications in the 
topology of neighboring exterior cells, some of which become triangles and 
pentagons. For instance, points $1$ and $2$ are displaced to boundary 
locations $a$ and $b$, forming a triangle with the background grid node $3$. 
Similarly, background node $10$ splits into points $e$ and $f$, producing a 
pentagonal cell $e$--$9$--$8$--$11$--$f$. Such pentagonal cells are further 
subdivided into a triangle and a quadrilateral, in this case $e$--$9$--$8$ and 
$e$--$8$--$11$--$f$.

\begin{figure}[h]
\begin{center}
\includegraphics[width=0.5\textwidth]{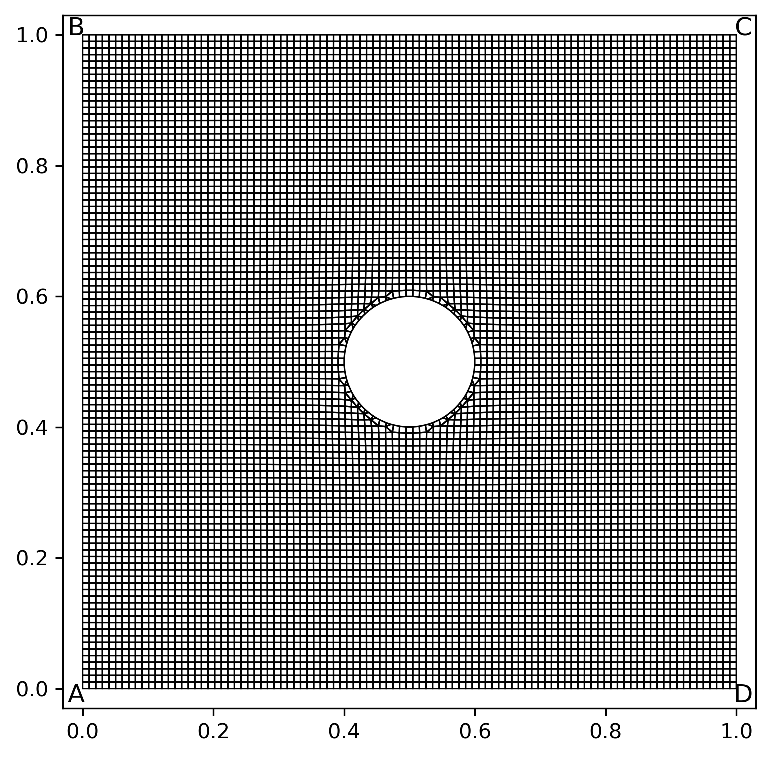}
\end{center}
\caption{The final mesh developed in a rectangular domain containing one immersed cylinder}
\label{fig:2}
\end{figure}

This procedure yields a set of boundary points on the immersed geometry that are, in general, non-uniformly spaced. To ensure adequate resolution and numerical robustness, these points are subsequently re-distributed to satisfy a prescribed minimum spacing along the boundary. A representative mesh obtained after this final adjustment for a geometry containing a single immersed cylinder is shown in Figure~\ref{fig:2}.

\section{Matrix-Free Potential-Flow Solver}

We employ the Finite Element Method (FEM), which provides excellent flexibility for handling mixed and irregular cell topologies, such as those generated in the present grid system.

\subsection*{Governing Equation}

For incompressible and irrotational flow, the velocity field can be expressed in terms of the stream function $\psi$ as
\[
\boldsymbol{u} = \nabla^\perp \psi =
\left(
\frac{\partial \psi}{\partial y},
-\frac{\partial \psi}{\partial x}
\right),
\qquad 
\nabla \cdot \boldsymbol{u} = 0,
\]
which directly leads to the Laplace equation
\begin{equation}
\nabla^2 \psi = 0. \label{LaPsi}
\end{equation}
This formulation can also be expressed in terms of the velocity potential
$\phi$, which satisfies the Laplace equation.
\begin{equation}
\nabla^2 \phi=0.
\end{equation}
 In this representation, the
velocity field is given by $\boldsymbol{u} = \nabla \phi$. The contour lines
of the potential $\phi$ and the stream function $\psi$ constitute an
orthogonal family of curves, forming the classical flow net.
\subsection*{Boundary Conditions}

Each immersed body $\Gamma_i$ is modeled as a constant-streamline boundary,
\begin{equation}
\psi\big|_{\Gamma_i} = C_i,
\end{equation}
where $C_i$ is the (unknown) constant value of the stream function on the $i$-th body. Prescribed normal velocities are accommodated by imposing the corresponding tangential derivative of the stream function, ensuring consistency with the potential-flow formulation. A reference streamline is additionally specified to eliminate the arbitrary additive constant inherent to the definition of $\psi$.

For bodies exhibiting corners or sharp variations in slope, special handling is required—most notably at trailing edges—to satisfy the Kutta condition. This is achieved by introducing an additional point in the fluid region, positioned such that the local flow remains tangential to the line connecting the trailing edge and the selected point. When necessary, a nearby fluid-region node may also be repositioned to ensure proper geometric alignment. The resulting configuration, combined with boundary nodes constrained to fixed values of the stream function, ensures consistent and robust enforcement of the trailing-edge condition.
\subsection*{Weak formulation}
The weak form of (\ref{LaPsi}), obtained through the standard variational
procedure (integration by parts omitted for brevity), is
\[
\int_{\Omega} \nabla\psi \cdot \nabla\eta \, d\Omega = 0,
\]
for all admissible test functions $\eta$. Using the Galerkin choice
$\eta = N$, with $N$ the row vector of element shape functions, the
element-level finite element relation becomes
\begin{equation}
\mathbf{K}_e\,[\psi]_e = \mathbf{f}_e,
\qquad
\mathbf{K}_e = \int_{\Omega_e} (\nabla N)^T (\nabla N)\, d\Omega.
\end{equation}

The right-hand side $\mathbf{f}_e$ includes contributions from Neumann
conditions on $\Gamma_{e,N}$,
\[
\mathbf{f}_e^{(N)} = \int_{\Gamma_{e,N}} N^{T}\,\bar{q}\, d\Gamma,
\]
and the adjustments arising from enforcing Dirichlet (essential) boundary
conditions through the adopted constraint-enforcement procedure. 

\subsection*{Matrix-Free Conjugate Gradient with Multiple Constraint Enforcement}

In the present implementation, the global stiffness matrix is not assembled.
Instead, element stiffness matrices are computed once and stored, and all
matrix--vector products required by the Conjugate Gradient (CG) method are
performed in a matrix-free manner. At each CG iteration, the contribution of
each element is obtained by multiplying the corresponding element stiffness
matrix with the element extract of the global vector, and the resulting
element-level vectors are accumulated into the global vector structure. This
significantly reduces memory requirements and allows computations with very
large numbers of degrees of freedom.

Multiple constraints are enforced directly at the vector level. All Dirichlet
boundary conditions—including  stream-function values on immersed
bodies and the Kutta condition—are imposed using a constraint-value enforcement
procedure applied immediately after each matrix--vector operation and solution
update. Because each node carries a single degree of freedom, the enforcement
is applied directly to those vector entries corresponding to constrained nodes.
Only the element equations associated with boundary nodes therefore become
non-homogeneous; interior elements retain homogeneous relations.

Let $N_{\mathrm{int}}$ denote the number of interior nodes, $N_{\mathrm{N}}$ the
number of nodes where Neumann or natural boundary conditions are applied, and
$N_{\mathrm{b}}$ the number of immersed solid bodies, each contributing a single
Dirichlet constraint (including Kutta conditions where applicable). The total
number of active unknowns is
\[
N_{\mathrm{dof}}
= N_{\mathrm{int}}
+ N_{\mathrm{N}}
+ N_{\mathrm{b}}.
\]

The algorithm for the matrix-free implementation of the Conjugate Gradient solver, including the enforcement of the multi-point constraint for the stream function on the immersed solid surface, is provided in \ref{app:algo}. The Conjugate Gradient iterations are advanced until the solution meets a prescribed convergence criterion. Specifically, the algorithm terminates when the global residual norm $\|r^{k}\|_{2}$---computed as the Euclidean norm of the assembled residual vector at iteration $k$---falls below a user-defined tolerance of $\texttt{tol}=10^{-12}$. To ensure robustness in situations where convergence may slow down or stall, the iterations are also limited by a maximum number of steps, denoted $\texttt{k\_max}$. Both \texttt{tol} and \texttt{k\_max} appear explicitly in the pseudocode and govern the stopping condition, ensuring that the solver remains both accurate and computationally bounded.

\subsection*{Velocity Field and Surface Tangential Velocity}

Once the stream function $\psi$ is known, the velocity field follows from
\[
u = \frac{\partial \psi}{\partial y},
\qquad
v = -\,\frac{\partial \psi}{\partial x}.
\]
Let $\boldsymbol{n}$ denote the outward unit normal on the body surface
$\Gamma_i$. The magnitude of the surface tangential velocity is
\[
u_t = \, \nabla \psi \cdot \boldsymbol{n} \,.
\]

In the present discretization, $u_t$ is evaluated using a finite–difference
approximation of the normal derivative of $\psi$ at the boundary. Let
$\psi_b$ be the evaluated stream-function value on the body surface,
$\psi_c$ the average stream-function value of the adjacent fluid element, and
$d_{nc}$ the perpendicular distance from the element centroid to the
boundary edge along the outward normal. The discrete tangential velocity is
then computed as
\[
u_t \;\approx\;
\, \frac{\psi_c - \psi_b}{d_{nc}} \,,
\]
which provides a consistent approximation of the normal derivative
$\partial \psi/\partial n$ and hence of the surface tangential velocity.
\section{Validation}
\begin{figure}[h]
\begin{center}
\includegraphics[width=\textwidth]{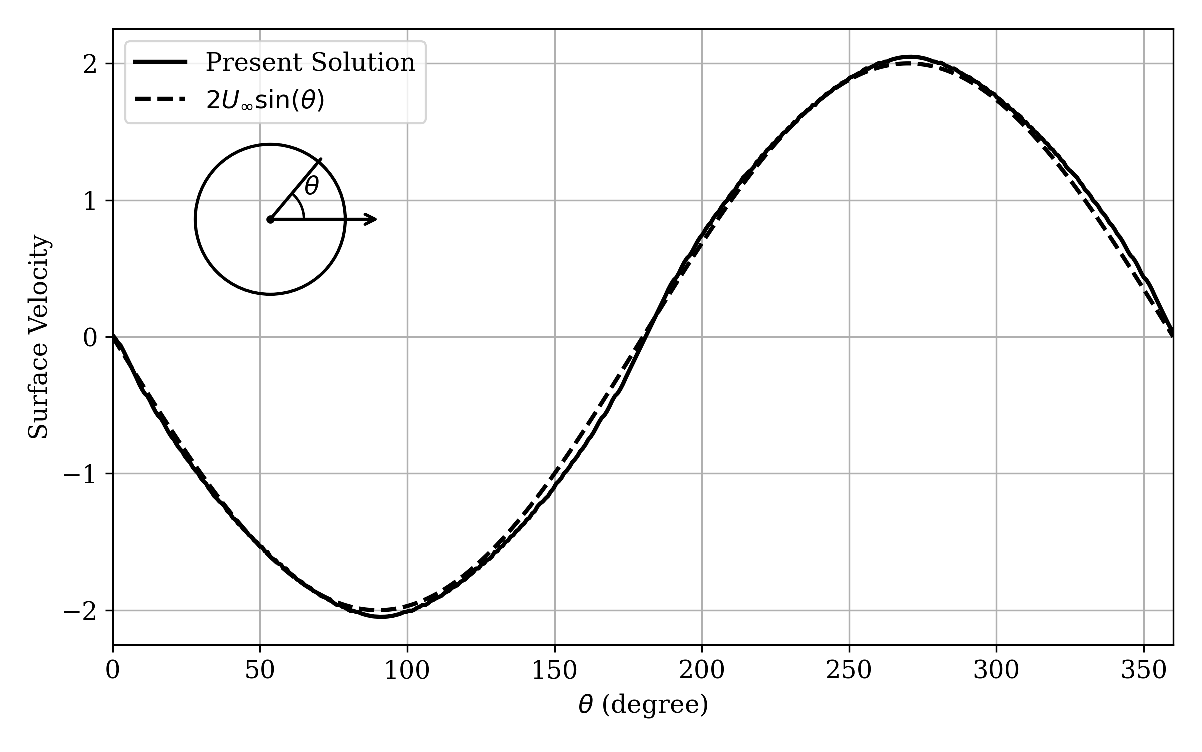}
\end{center}
\caption{Surface velocity distribution over a cylinder and comparison with the analytical solution}
\label{fig:2-r}
\end{figure}

The potential flow past a cylinder confined between two parallel walls is computed using the developed numerical solver to analyze the resulting surface--velocity distribution. The geometric configuration---including the domain size, cylinder location, and wall spacing---is illustrated in the mesh diagram shown in Figure~\ref{fig:2}. The bottom (AD) and top (BC) boundaries are prescribed as streamlines by imposing Dirichlet boundary conditions on the stream function, with $\psi = 0$ and $\psi = 1~m^2/s$, respectively. The natural boundary conditions applied on AB and CD enforce zero tangential velocity, thereby causing one of these boundaries to act as the inlet and the other as the outlet. The prescribed variation of $\psi$ between the walls corresponds to a uniform inlet and outlet velocity of $1~m/s$.
The computed surface--velocity distribution is shown in Figure~\ref{fig:2-r}, together with the analytical solution for potential flow past an isolated cylinder. The two distributions exhibit excellent agreement, with the present computation showing only a slight overprediction and a maximum percentage error of approximately $2.1\%$. This small deviation is primarily attributed to the confinement effect imposed by the streamline boundaries AD and BC. The predicted value of the stream function at the immersed cylinder surface is $0.50001959$, which deviates only negligibly from the exact value of $0.5$, thereby clearly validating the multi-point constraint implementation.

\section{Results and Discussion}
The computational tools developed in this work are applied to solve the potential flow past multiple immersed bodies, with the objective of understanding the interaction effects, flow blockage in the throat regions and the consequent surface tangential velocity changes.

\subsection*{Potential Flow Interfences in a Multiple--Body System}

\begin{figure}[h]
\begin{center}
\includegraphics[width=\textwidth]{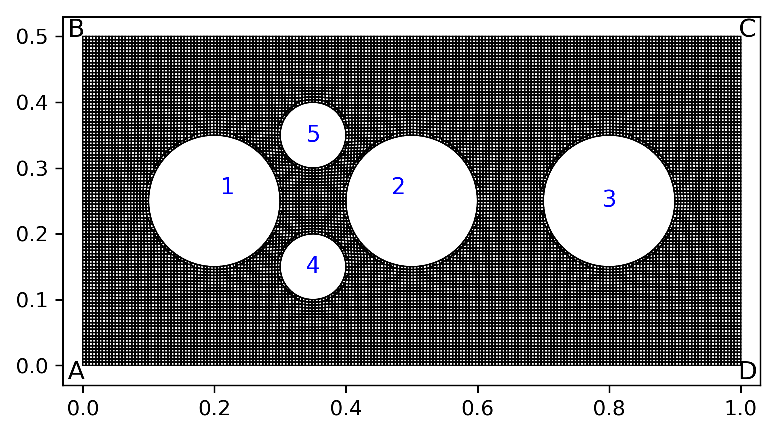}
\end{center}
\caption{Mesh in a geometry consisting of five cylinders}
\label{fig:3}
\end{figure}

\begin{figure}[h]
\begin{center}
\includegraphics[width=0.55\textwidth]{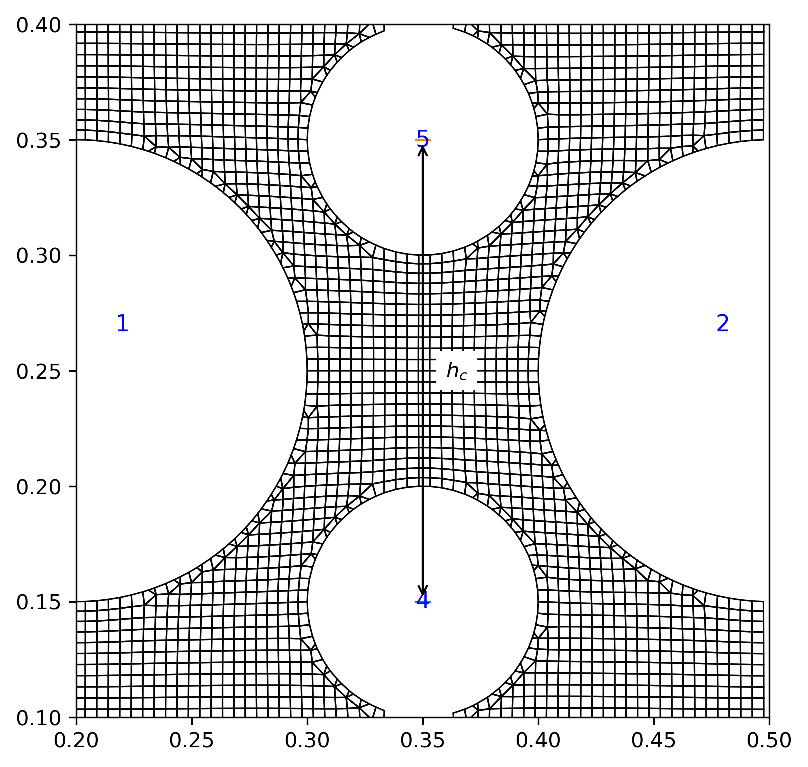}
\end{center}
\caption{Zoomed view of the grid generated for the five-cylinder immersed system, showing details around cylinders 1, 2, 4, and 5.}
\label{fig:4}
\end{figure}

The configuration analysed consists of five circular solids immersed in a channel flow between two parallel plates. Three cylinders, numbered~1, 2, and~3, are aligned along the \(x\)-axis, while the remaining two cylinders, 4 and~5, are named transverse cylinders, are positioned in the transverse (\(y\)) direction. To investigate the potential interaction effects, we vary the ratio of the distance between transverse cylinders, marked as \(h_c\) in figure (\ref{fig:4}), to the height of the computational domain \(h_d\), marked as AB in figure (\ref{fig:3}). This nondimensional parameter called gap ratio is defined as \(\delta = h_c / h_d\) and its effect on the flow pattern and velocity variations is examined in this study. The minimum value of \(\delta\) of this flow configuration \(\delta_{min}=2D\), where \(D\) is the diameter of the transverse cylinder.

Figure~\ref{fig:3} shows the mesh generated around the immersed bodies as well as the principal geometric dimensions of the configuration. The computational domain is
\(
D = \{(x,y)\mid x \in [0,1~\mathrm{m}],\; y \in [0,0.5~\mathrm{m}]\}.
\)
The construction of the finite elements within this domain, particularly in the regions surrounding the immersed bodies, is illustrated in the zoomed-in mesh view shown in Figure~\ref{fig:4}.

\begin{figure}[h]
\begin{center}
\includegraphics[width=\textwidth]{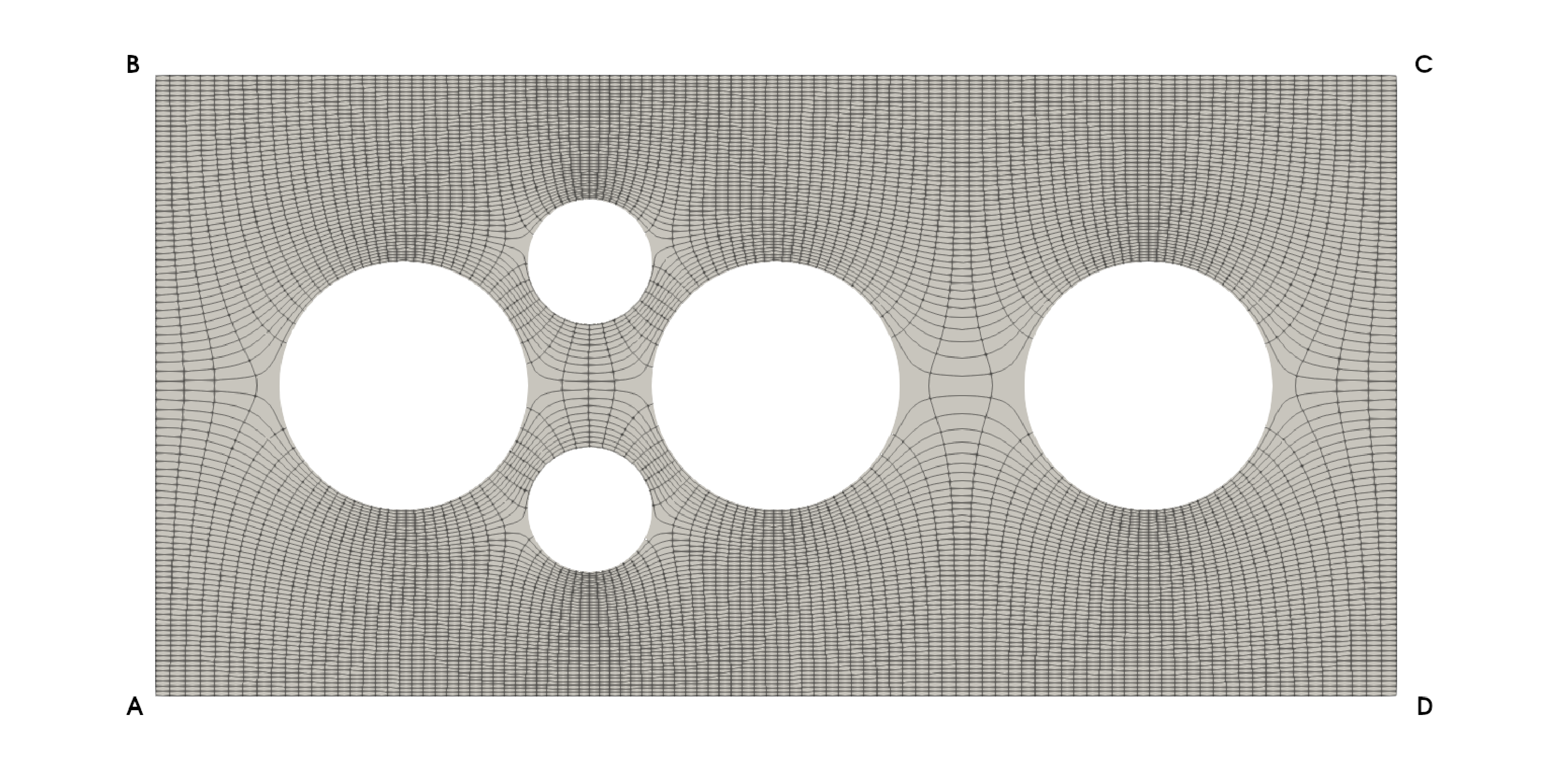}
\end{center}
\caption{Flow-net for the multi-cylinder configuration with gap ratio  \(\delta=0.4\).}
\label{fig:5}
\end{figure}

\begin{figure}[h]
\begin{center}
\includegraphics[width=\textwidth]{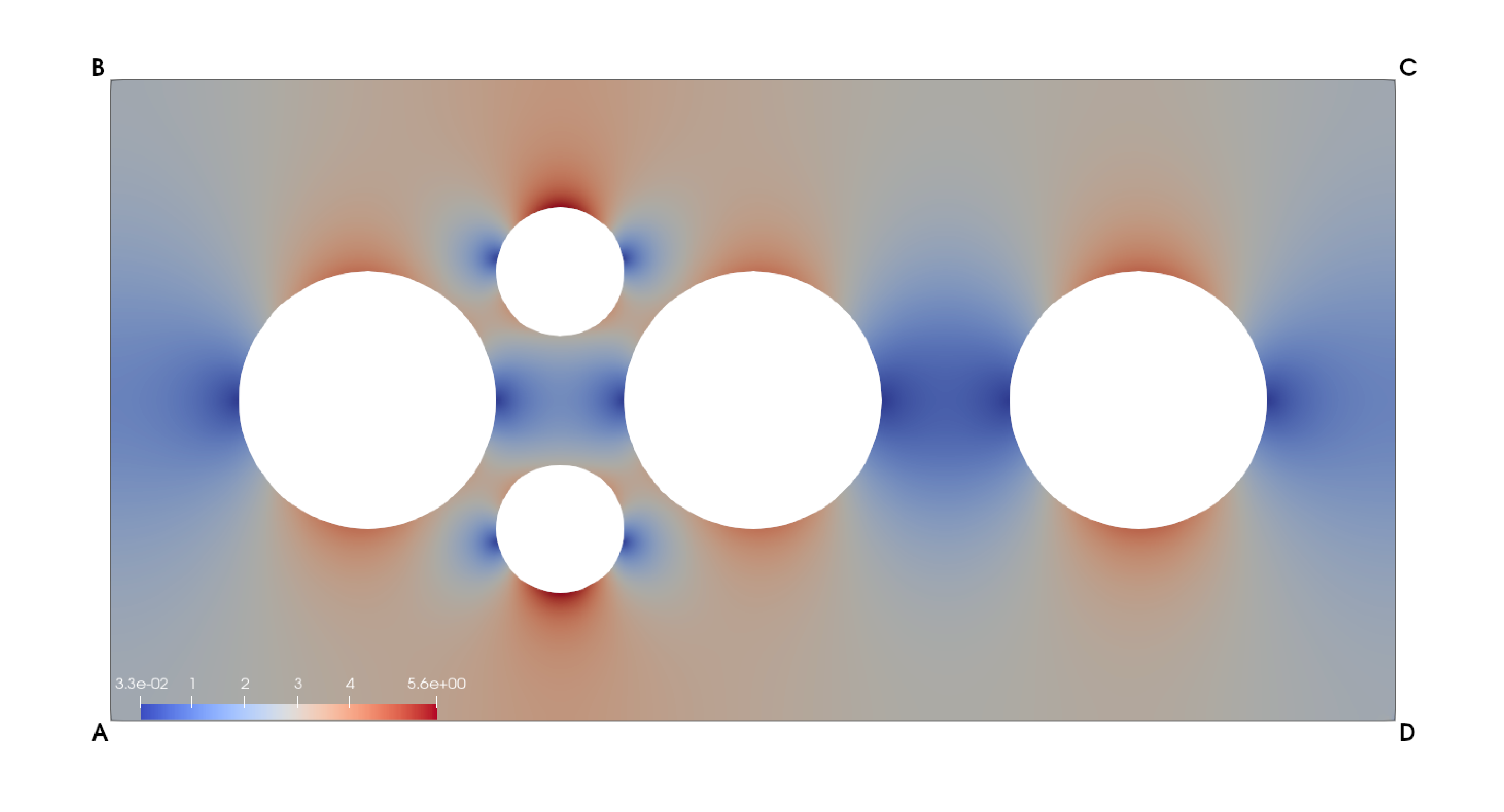}
\end{center}
\caption{Color-filled band plot of the velocity magnitude in the five-cylinder configuration for $\delta = 0.4$.
}
\label{fig:6}
\end{figure}

\begin{figure}[h]
\begin{center}
\includegraphics[width=\textwidth]{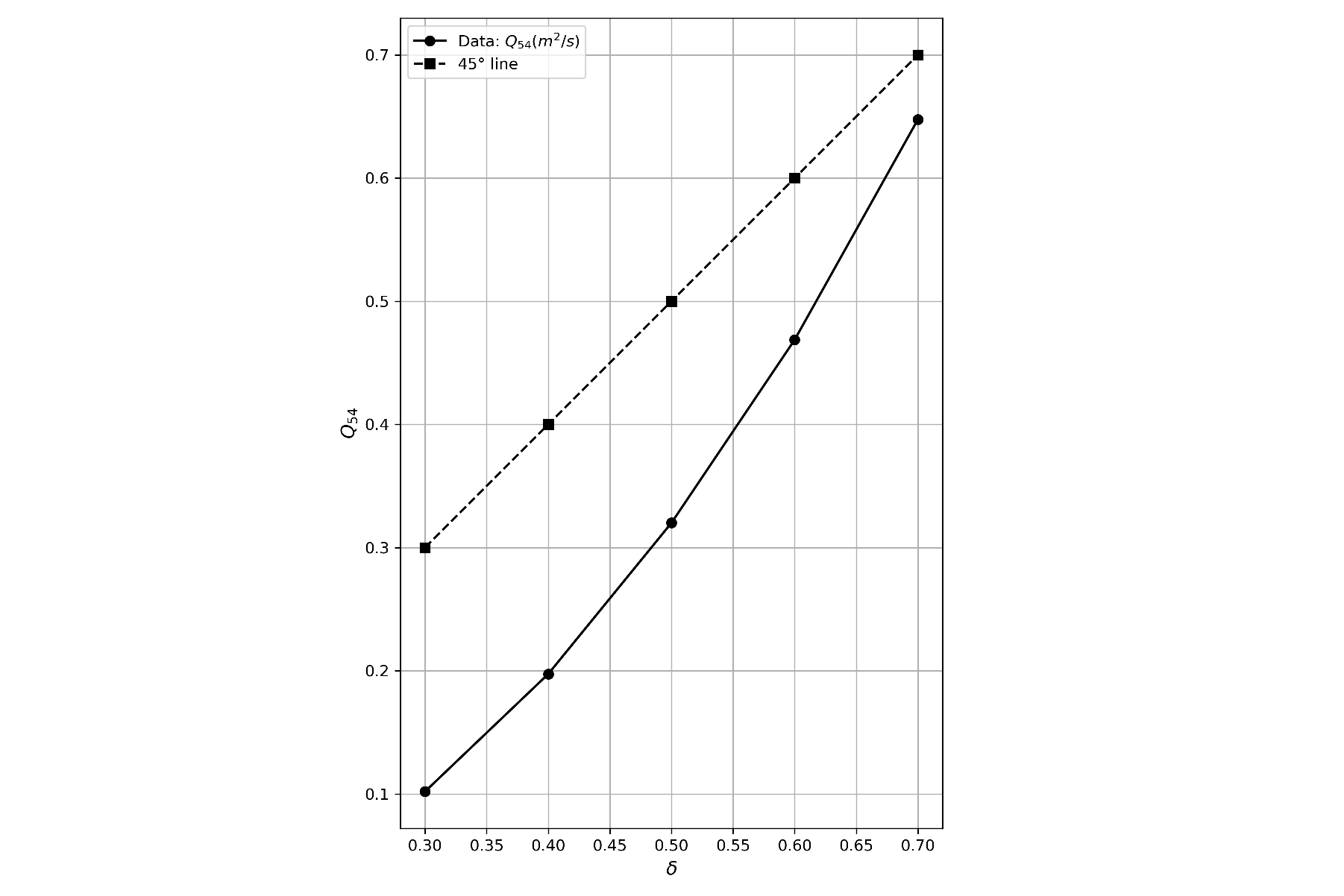}
\end{center}
\caption{Variation of the flow rate per unit width in between the transverse cylinders, 4 and 5 as a function of the gap ratio \(\delta\).}
\label{fig:7}
\end{figure}

The diameters of cylinders~1, 2, and~3 are each \(0.2\,\mathrm{m}\), while cylinders~4 and~5 have diameters of \(0.1\,\mathrm{m}\), and so \(\delta_{min}=0.2\). Boundaries~AD and~BC are treated as streamlines, representing parallel plates with prescribed stream-function values of \(0\) and \(1\,\mathrm{m}^2/\mathrm{s}\), respectively, which correspond to a uniform inflow velocity of \(2\,\mathrm{m/s}\).

The results confirm that no explicit enforcement of the Kutta condition is required to determine the rear stagnation point, consistent with classical potential-flow theory~\cite{katz2001low,trefethen1980potential}. Figures~\ref{fig:5} and~\ref{fig:6} present the flow net and the velocity-magnitude field for gap ratio \(\delta = 0.4\). The velocity magnitudes attain locally higher values near the top and bottom surfaces of the cylinders, whereas the upstream and downstream regions exhibit local minima. The transverse cylinders push the flow toward the gap between cylinders~1 and~2; however, because there is no artificial turning of the flow from the top and bottom surfaces of cylinder~2, towards cylinder-3, the flow velocity reaches a global minimum in the gap between the two cylinders. The turning of the streamlines and the differences noted in the gap between stream lines  in these regions, visible in Figure~\ref{fig:5}, clearly demonstrates this effect.

\subsection*{Effect of Gap Ratio \(\delta\)}

The gap ratio is varied as \(\delta = 0.3,\,0.4,\,0.5,\,0.6,\) and \(0.7\) in this study. Table~\ref{Tab:1} presents the predicted stream-function values over cylinder~1 and over the transverse cylinders~4 and~5. The total imposed flow rate per unit width is \(1~\mathrm{m}^2/\mathrm{s}\). Owing to symmetry, the horizontally aligned cylinder~1 divides the flow equally into the upper and lower regions, regardless of the value of \(\delta\), and its stream-function value therefore remains fixed at \(0.5~\mathrm{m}^2/\mathrm{s}\). The flow rate per unit width between the transverse cylinders is obtained from the difference in their stream-function values, denoted as \(Q_{54} = \psi_5 - \psi_4\). In a channel containing only the two transverse cylinders (i.e., without the horizontally aligned cylinders), for \(\delta>\delta_{min}=0.2\), the gap ratio \(\delta\) itself represents the flow rate per unit width between cylinders~4 and~5. Figure~\ref{fig:7} illustrates the variation of \(Q_{54}\) with \(\delta\), along with the corresponding uninfluenced flow represented by a \(45^\circ\) line. The vertical difference between these curves indicates the reduction in flow rate caused by irrotational interactions, and this reduction diminishes as \(\delta\) increases.

To quantify this effect, the ratio of the actual flow rate \(Q_{54}\) (with interference from the horizontally aligned cylinders) to the corresponding value without interference (\(\delta\)) is defined as the \textit{Coefficient of Potential Flow Interaction} (CoPFI) is introduced
\[
\text{CoPFI} = \frac{Q_{54}}{\delta}.
\]

\begin{figure}[h!]
    \centering

    \begin{subfigure}{0.45\textwidth}
        \centering
        \includegraphics[width=\textwidth]{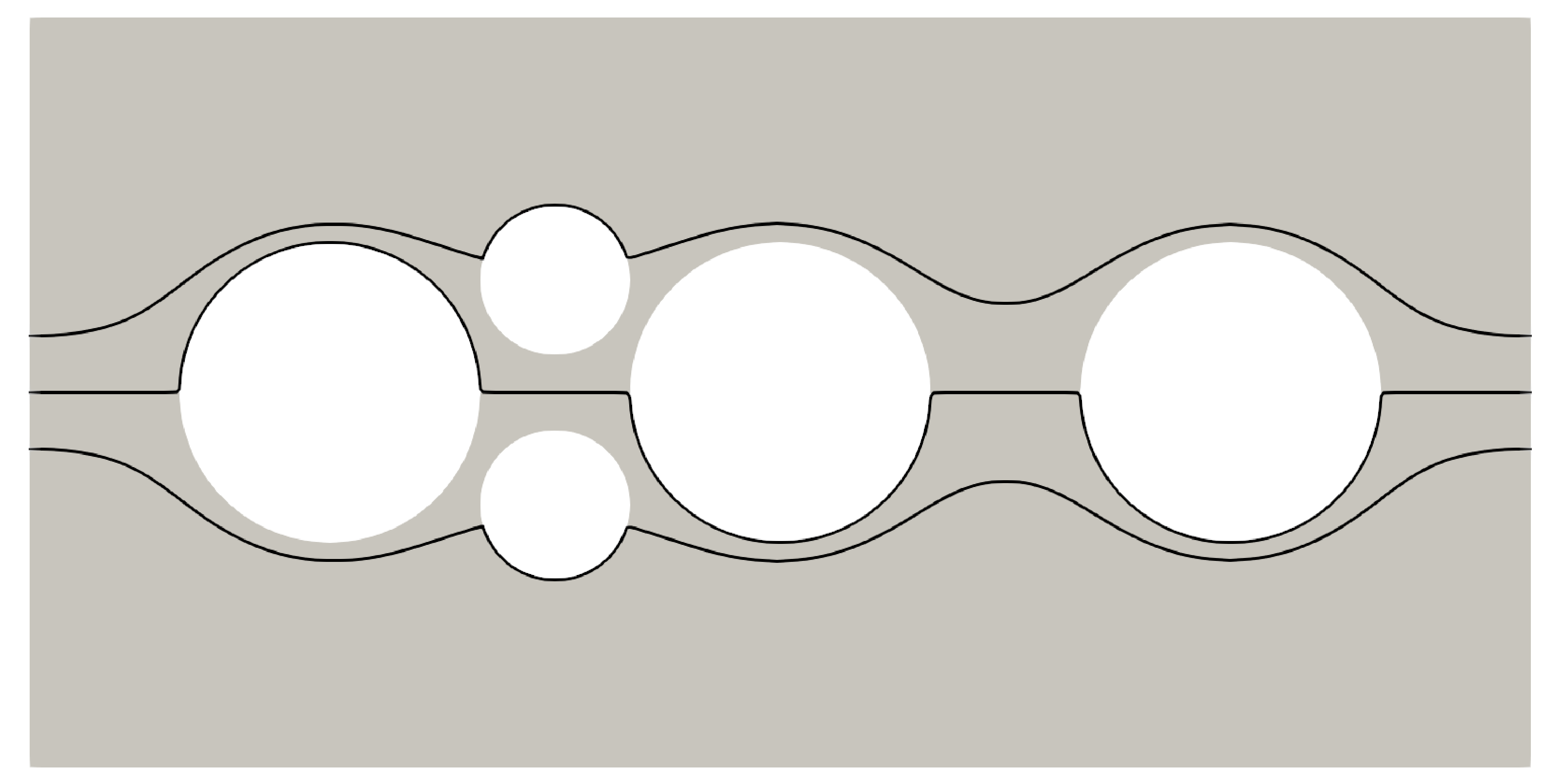}
        \caption{\(\delta=0.3\)}
        \label{fig:row1a}
    \end{subfigure}
    \hfill
    \begin{subfigure}{0.45\textwidth}
        \centering
        \includegraphics[width=\textwidth]{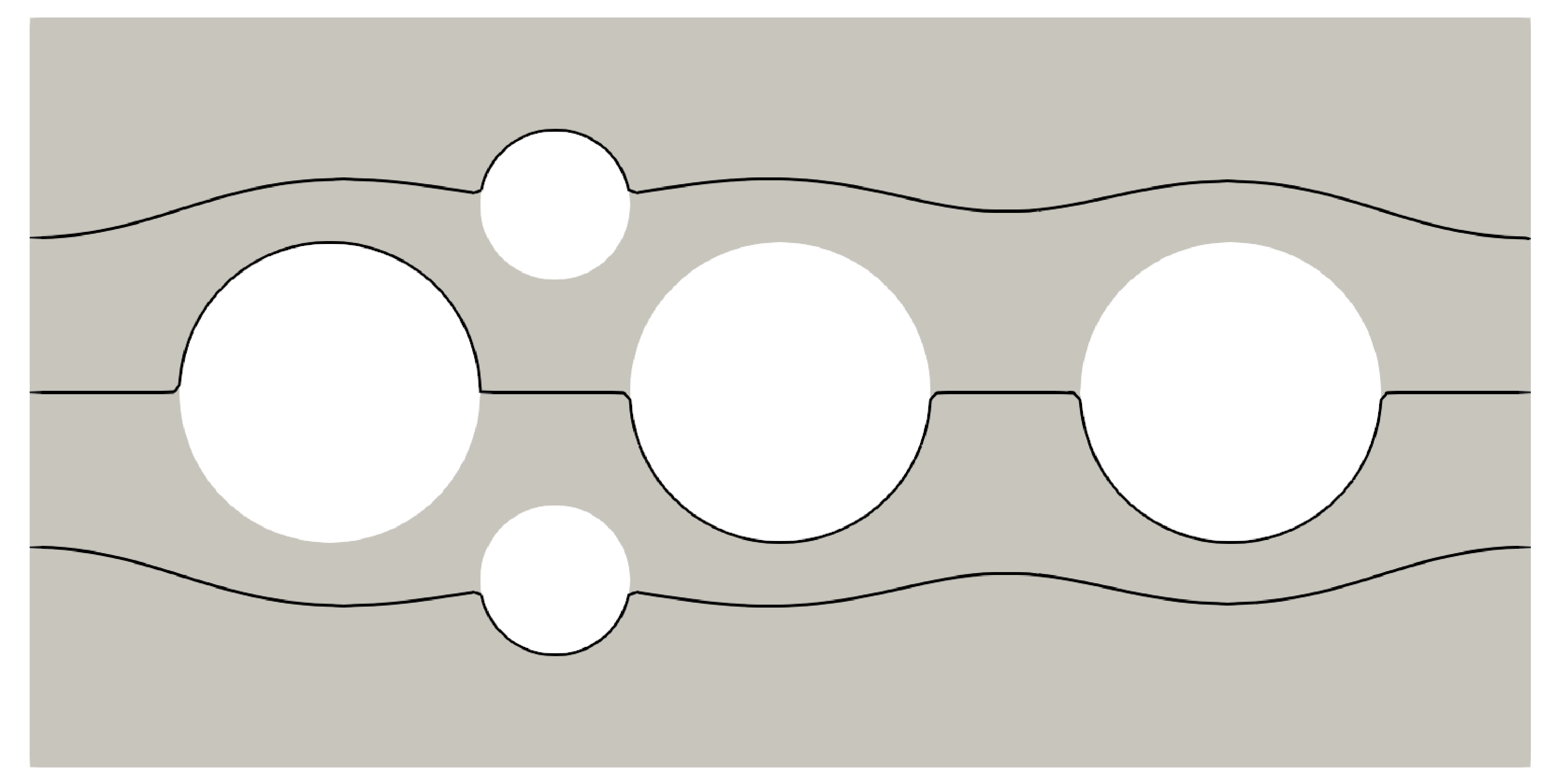}
        \caption{\(\delta=0.5\)}
        \label{fig:row1b}
    \end{subfigure}

    \vspace{0.6cm}

    \begin{subfigure}{0.45\textwidth}
        \centering
        \includegraphics[width=\textwidth]{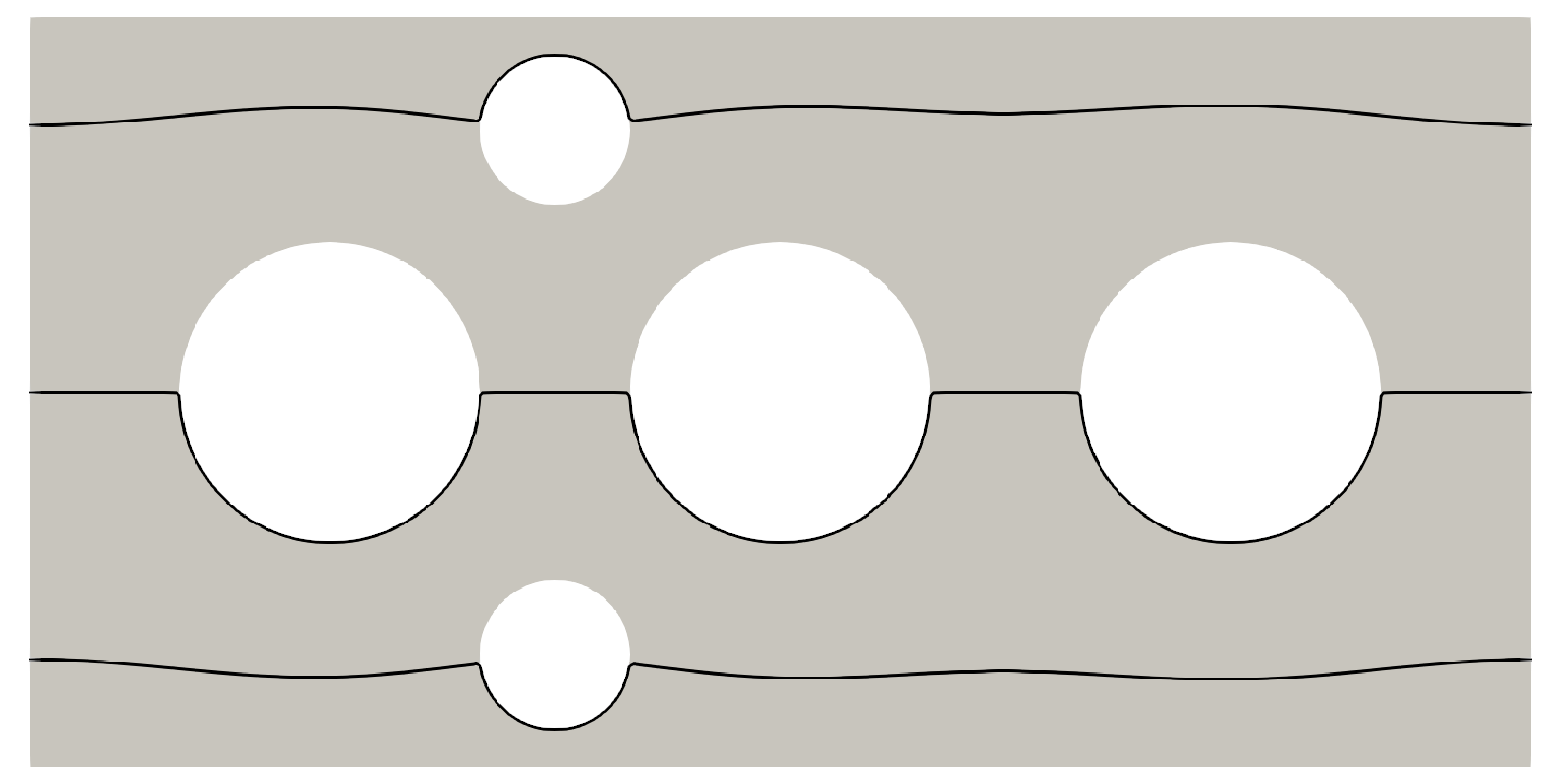}
        \caption{\(\delta=0.7\)}
        \label{fig:row2}
    \end{subfigure}

    \caption{Plot of the stagnation streamlines between cylinders 4 and 5 and those enclosing the in-line cylinders 1, 2, and 3.}
    \label{fig:st-combined}
\end{figure}

This coefficient measures the extent to which the horizontally aligned cylinders modify the flow between the transverse cylinders. As shown in Table~\ref{Tab:1}, the decrease in CoPFI with decreasing \(\delta\) reflects the strengthening of the \emph{mutual irrotational interaction} between the two sets of cylinders at smaller gap ratios. A cubic polynomial fit for \(\mathrm{CoPFI}\) as a function of \(\delta\), with an \(R^2\) value very close to unity, and a Mean Square Error (MSE) of \(1.40428920e-07\) is given by
\[
\mathrm{CoPFI}(\delta)
= 0.80417\,\delta^{3}
- 1.38561\,\delta^{2}
+ 2.21409\,\delta
- 0.22171 .
\]
The planar streamtubes formed between the stagnation streamlines of the transverse cylinders, together with the corresponding evolution of the stagnation streamlines  of cylinders~4 and~5 for $\delta = 0.3,\,0.5,$ and $0.7$, are presented in Figure~\ref{fig:st-combined}. As $\delta$ increases, these enclosing stagnation envelopes become progressively smoother, reflecting a reduction in potential-flow interference. Conversely, at lower gap ratios ($\delta$), the deformation-wavyness of the stagnation envelope is more pronounced, signifying higher levels of potential-flow interaction—consistent with the observed reduction in the coefficient of potential-flow interference (CoPFI) at smaller $\delta$.

\begin{figure}    
    \begin{subfigure}{0.45\textwidth}
        \centering
        \includegraphics[width=\linewidth]{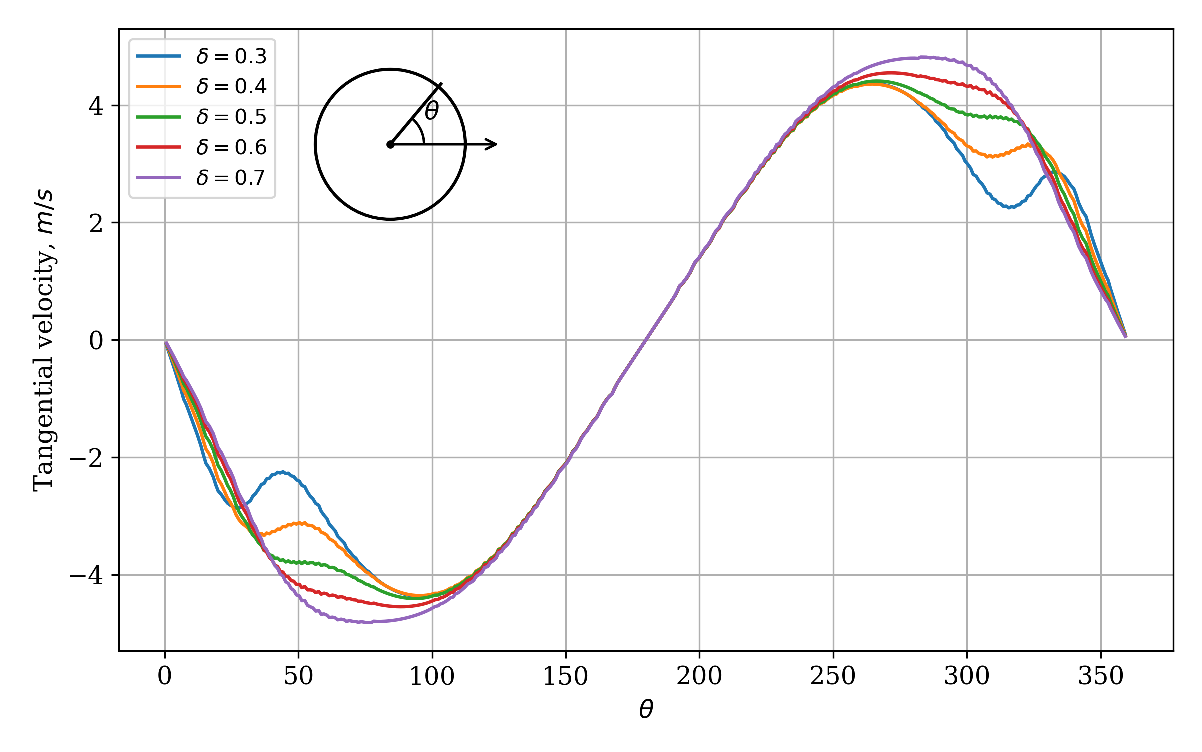}
        \caption{Cylinder-1}
    \end{subfigure}
    \hfill
    \begin{subfigure}{0.45\textwidth}
        \centering
        \includegraphics[width=\linewidth]{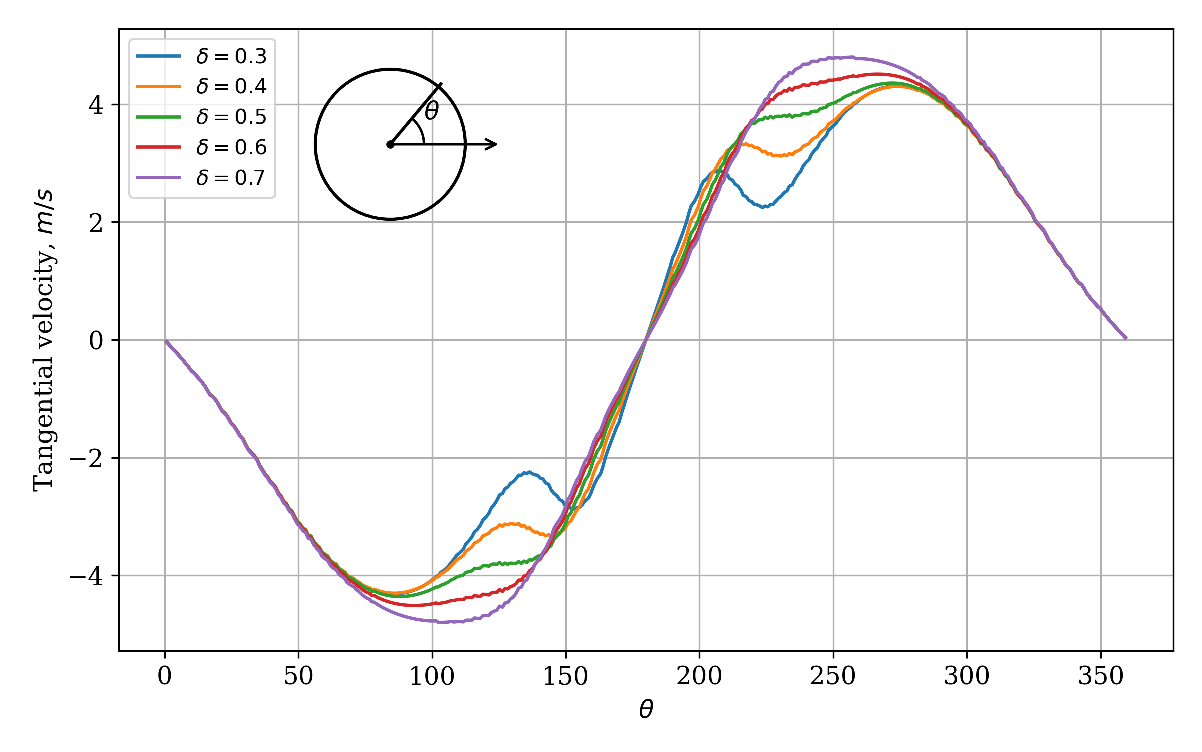}
        \caption{Cylinder -2 }
    \end{subfigure}

    \vspace{0.3cm}

    \begin{subfigure}{0.45\textwidth}
        \centering
        \includegraphics[width=\linewidth]{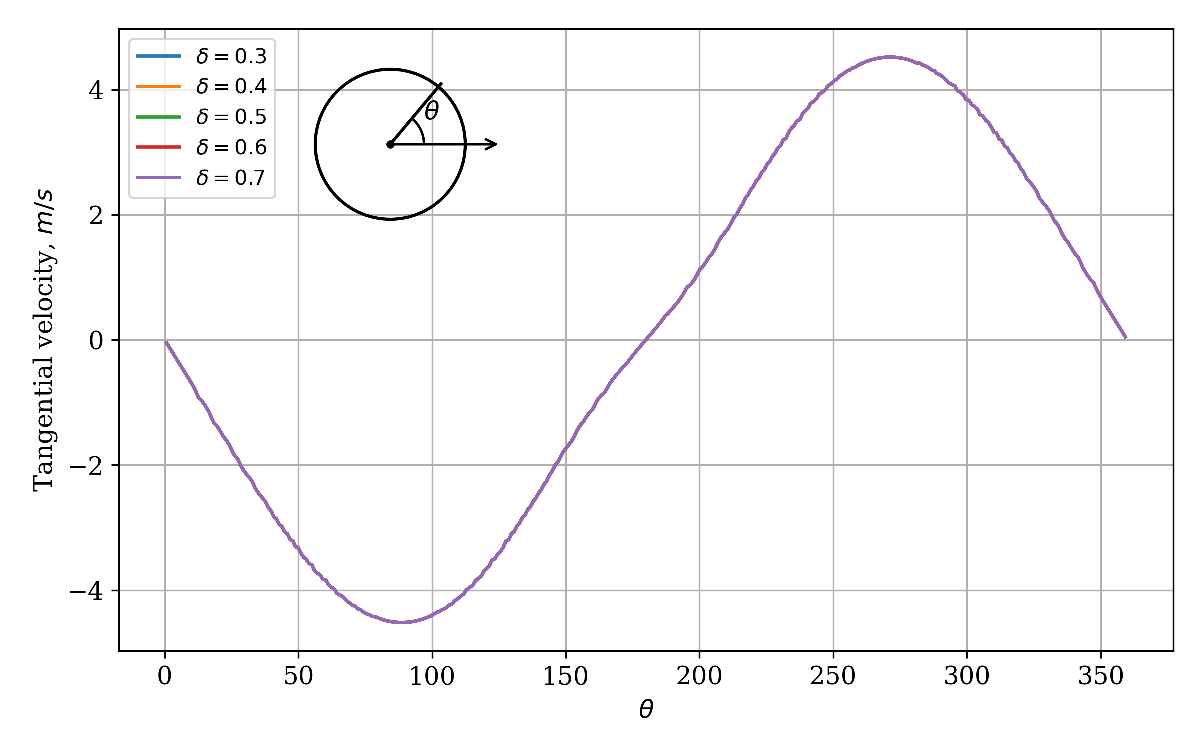}
        \caption{Cylinder -3}
    \end{subfigure}
    \hfill
    \begin{subfigure}{0.45\textwidth}
        \centering
        \includegraphics[width=\linewidth]{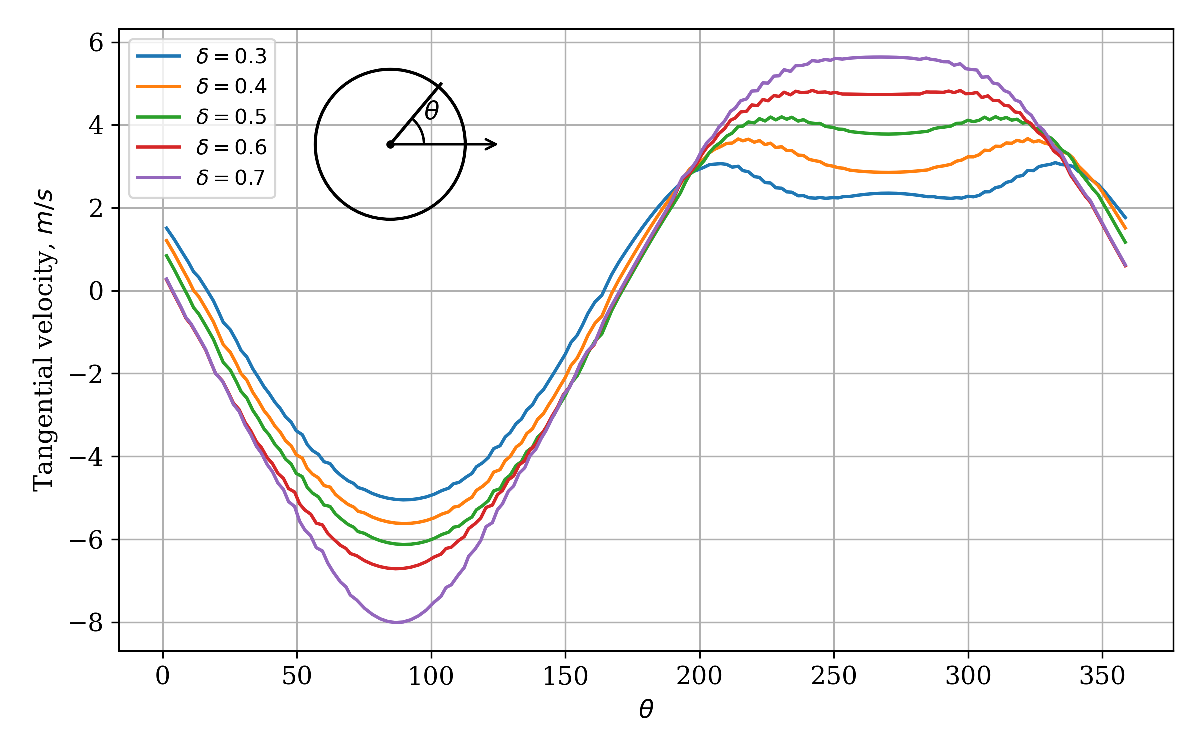}
        \caption{Cylinder -5}
    \end{subfigure}

    \caption{Tangential velocity profiles for different values of $\delta$ on the surface of cylinders 1,2,3, and 5.}
    \label{fig:cyl-all}
\end{figure}

Although the magnitude of potential-flow interaction is quantified through the CoPFI, its influence on the tangential surface-velocity distribution over the strongly interacting cylinders~1, 2, 4, and~5, as well as the weakly interacting cylinder~3, is also examined. Because cylinders~4 and~5 experience identical symmetric interference, the tangential velocity distributions for cylinders~1, 2, 3, and~5 as functions of the circumferential angle $\theta$ are shown in Figure~\ref{fig:cyl-all}. The results indicate that the downstream region of cylinder~1 for $\theta \in [0,100]\cup[270,360]$ and the upstream region of cylinder~2 for $\theta \in [90,270]$ exhibit wavy variations, the intensity of which increases as $\delta$ decreases. These oscillations are attributed to enhanced potential-flow interaction at small gap ratios. A similar wavy behaviour is observed on the lower half of cylinder~5 for $\theta \in [180,360]$, and, although not separately illustrated, an identical trend occurs on the corresponding upper half of cylinder~4 for $\theta \in [0,180]$.

Across all interacting cylinders, the presence of potential-flow interference leads to a reduction in the peak surface velocity, and the magnitude of this reduction diminishes with increasing $\delta$. In contrast, the velocity distribution on cylinder~3—located farther from the transverse interacting cylinders—remains essentially unaffected by variations in $\delta$, clearly demonstrating that potential-flow disturbances dissipate over short distances within the flow field.

\begin{table}
\caption{Variation of stream-function values over cylinders~1, 4, and~5; the volume flow rate per unit width $Q_{54}~\mathrm{m}^2/\mathrm{s}$; and the Coefficient of Potential Flow Interaction (CoPFI). The total imposed flow rate is $Q = 1~\mathrm{m}^2/\mathrm{s}$.}
\label{Tab:1}
\centering
\small
\renewcommand{\arraystretch}{1.5}
\begin{tabular}{l l l l l l}
\hline\hline
$\delta$ & $\psi_1$ & $\psi_5$ & $\psi_4$ & $Q_{54}=\psi_5-\psi_4$ & CoPFI = $Q_{54}/\delta$ \\
\hline
0.3 & 0.50000 & 0.55094 & 0.44905 & 0.10189 & 0.33963 \\
0.4 & 0.50001 & 0.59866 & 0.40134 & 0.19732 & 0.49330 \\
0.5 & 0.50001 & 0.66001 & 0.33999 & 0.32003 & 0.64006 \\
0.6 & 0.50001 & 0.73437 & 0.26563 & 0.46874 & 0.78123 \\
0.7 & 0.50000 & 0.82380 & 0.17620 & 0.64760 & 0.92514 \\
\hline\hline
\end{tabular}
\normalsize
\end{table}

\section{Conclusions}

This work presented an integrated computational framework for potential-flow analysis around multiple immersed bodies, combining a simple body-conforming mesh-generation procedure, a matrix-free finite-element solver for the Laplace equations governing $\psi$ and $\phi$, and a multi-point constraint strategy for determining the stream-function values associated with all immersed solids. The methodology was applied to configurations involving five cylinders placed both horizontally and transversely within a channel, enabling detailed evaluation of inter-body irrotational interactions.

The results demonstrate that the proposed grid-generation technique provides well-resolved body-fitted meshes for geometries containing multiple immersed solids, while the matrix-free finite-element formulation yields accurate solutions with significantly reduced memory requirements. The computed flow nets and velocity-magnitude fields reproduce classical potential-flow behaviour, including the solution-determined stagnation-point locations without the need for enforcing a Kutta-type condition. These results confirm the solver's robustness in capturing the essential features of irrotational flow around multiple obstacles.

A systematic assessment of potential-flow interference was carried out by varying the gap ratio $\delta$ between the transverse cylinders. The flow rate between the cylinders, expressed through $Q_{54}$, decreases substantially at small $\delta$ due to strong irrotational interactions induced by the horizontally aligned cylinders. The Coefficient of Potential Flow Interaction (CoPFI), provides a quantitative measure of this effect and exhibits a rapidly decreasing trend as $\delta$ approaches its minimum geometric limit. A cubic polynomial expression  for CoPFI as a function of $\delta$, is reported nabling predictive interpolation.

The streamline envelopes and planar streamtubes illustrate progressive smoothing of stagnation boundaries as $\delta$ increases, consistent with a weakening of potential-flow interference. Surface-velocity distributions on cylinders~1, 2, 4, and~5 reveal wavy variations whose amplitudes grow as $\delta$ decreases, indicating intensified inter-body interaction under tighter spacing. In contrast, cylinder~3, positioned away from the transverse pair, shows negligible sensitivity to $\delta$, demonstrating that irrotational disturbances dissipate rapidly over short distances.

Overall, the study demonstrates that the proposed computational framework is efficient, accurate, and well suited for multi-body potential-flow simulations, as well as for applications requiring repeated evaluations such as optimization or parametric design. The methodology establishes a solid foundation for future extensions, including shape optimization, automated sensitivity analysis, and coupling with approximate boundary-layer computations to incorporate viscous effects in the vicinity of solid surfaces.

\subsection{Data Availability Statement}
Some or all data, models, or code that support the findings of this study are available from the corresponding author upon reasonable request.
\label{section:references}
\bibliography{ascexmpl-new}

\newpage

\section{Appendixes}
\renewcommand{\thesection}{Appendix \Alph{section}}
\subsection{Algorithm} \label{app:algo}
\begin{algorithm}[h!]
\caption{CG solver (pseudocode) — faithful to the FORTRAN implementation}
\begin{algorithmic}[1]
\STATE \textbf{Input:} element matrices $\{A_e\}$, element RHS $b$, initial $x$
\FOR{each element $e$}
  \FOR{$i=1:\text{nne}_e$}
    \STATE $v_{i,e} \gets -\,b_{i,e}$
    \FOR{$j=1:\text{nne}_e$}
      \STATE $v_{i,e} \gets v_{i,e} + a_{i,j,e}\,x_{node(j,e)}$
    \ENDFOR
  \ENDFOR
\ENDFOR

\STATE Zero: $r\leftarrow 0,\;xp\leftarrow 0,\;rb\leftarrow 0,\;xpb\leftarrow 0$
\STATE Assemble $v_e$ into global residuals $r$ (interior nodes) and $rb$ (body DOFs)
\STATE Initialize $xp_i \leftarrow r_i$ for interior nodes; $xpb_k \leftarrow rb_k$
\STATE Copy $xpb_k$ to $xp$ at body boundary node indices

\STATE $r_{\text{norm}} \gets \sum_{i\in\text{int}} r_i^2 + \sum_{k} rb_k^2$
\STATE $k \gets 0$

\WHILE{ $r_{\text{norm}} > \text{tol}$ \AND $k < k_{\max}$ }
  \STATE $a_n \gets r_{\text{norm}}$
  \STATE \% Compute $Ap$: $v_e \gets -A_e\,xp_e$
  \FOR{each element $e$}
    \FOR{$i=1:\text{nne}_e$}
      \STATE $v_{i,e}\gets 0$
      \FOR{$j=1:\text{nne}_e$}
        \STATE $v_{i,e}\gets v_{i,e} - a_{i,j,e}\,xp_{node(j,e)}$
      \ENDFOR
    \ENDFOR
  \ENDFOR
  \STATE Assemble $v_e$ into $apval$ (interior) and $apvalb$ (body)
  \STATE $\alpha_d \gets \sum_{i\in\text{int}} apval_i\,xp_i + \sum_k apvalb_k\,xpb_k$
  \STATE $\alpha \gets a_n / (\alpha_d + \varepsilon)$
  \STATE Update $x, r, \text{psik}, rb$ and compute new $r_{\text{norm}}$
  \STATE $\beta \gets r_{\text{norm}} / (a_n + \varepsilon)$
  \STATE Update search vectors $xp \gets r + \beta xp$, $xpb \gets rb + \beta xpb$
  \STATE Copy $xpb$ into $xp$ at body boundary nodes
  \STATE $k \gets k+1$
\ENDWHILE

\STATE Copy final body values $\text{psik}_k$ into $x$ at body-node indices
\STATE \textbf{Output:} $x, r_{\text{norm}}$
\end{algorithmic}
\end{algorithm}

\end{document}